\documentstyle[aps,prb,twocolumn,psfig]{revtex}
\draft
\def\beq{\begin{equation}}
\def\eeq{\end{equation}}
\def\beqarr{\begin{eqnarray}}
\def\eeqarr{\end{eqnarray}}

\input epsf.tex

\begin{document}
\draft

\twocolumn[\hsize\textwidth\columnwidth\hsize\csname @twocolumnfalse\endcsname

\title{London equation studies of thin-film superconductors with a triangular antidot lattice}
\author{Sa-Lin Cheng$^1$, D. J. Priour Jr$^{1,2}$, H. A. Fertig$^1$}
\address{$^1$Department of Physics and Astronomy, University of Kentucky, Lexington, KY40506-0055\\
	 $^2$Center for Computational Sciences, University of Kentucky, Lexington, KY40506-0045}

\date{\today}

\maketitle

\begin{abstract}
We report on a study of vortex pinning in nanoscale antidot defect arrays
in the context of the London Theory.  Using a wire network model, we 
discretize the array with a fine mesh, thereby providing a detailed 
treatment of pinning phenomena.  The use of a fine grid has enabled
us to examine both circular and elongated defects, patterned in the 
form of a rhombus.  The latter display pinning characteristics superior 
to circular defects constructed with the similar area.  We calculate 
pinning potentials for defects containing zero and single quanta, and 
we obtain a pinning phase diagram for the second matching field, 
$H = 2 \Phi_{o}$. 
\end{abstract}


]

%
%
\section{Introduction}

   It has long been known that defects in superconductors can act
as pinning centers for vortices.  The trapping of vortices reduces 
dissipation and permits higher critical currents, which is important
in engineering applications involving superconductors.  In nature, 
one expects defects to be randomly distributed within a sample. To 
examine vortex pinning in a more controlled setting, a body of 
experimental work has examined pinning phenomena in artificial defect  
arrays. Advances in sub-micron lithography\cite{Metlushko,Gielen,Baert1,Baert2}
permit the construction of artificial nanoscale defect arrays. 
Lorentz microscopy permits direct imaging\cite{Harada} of the magnetic fields 
associated with the vortices and supercurrents. Scanning Hall probes\cite{Bell,Field} 
provide less detail, albeit over a greater spatial range. The defects can be fashioned 
as regions of excised superconductor, ``antidots'', or they may consist of ferromagnetic 
inclusions. This study concentrates on the antidot case.     

   The recent experimental work\cite{Harada,MoshB96,MoshB98,DeLong} on nanoscale 
defect arrays in thin films has motivated a number of theoretical 
studies\cite{Shapiro,Buzdin,Nori96,Nori98} of pinning phenomena in them. 
In this vein, we offer a detailed numerical treatment of thin-film periodic 
antidot arrays in the context of the London theory\cite{de Gennes}. 
Our study differs from most previous theoretical work in that it explicitly treats
the thin-film geometry, rather than a bulk system with two dimensional ordering.
Although we report on results obtained for the triangular antidot lattice, our method 
can readily be adapted to other geometries such as the square lattice. It is also within 
the scope of our method to consider more general inclusions, such as nanoscale defects 
with a permanent magnetic moment.  

   Since the London Theory does not admit a solution in closed form for our
system, a numerical approach is needed.  In our case, the numerical treatment
involves solving the London equations in discrete form.  In doing so, we model
the supercurrents in our system as a wire network. To satisfy the zero-current constraint 
within the inclusions, we expand the current in terms of elements of a special basis, 
to be described in greater detail in the next section. To determine the optimum current 
configuration, we minimize the discretized version of the London free energy with respect 
to the basis expansion coefficients.

\vspace{8mm}
\centerline{\bf Synopsis of the results}
\vspace{5mm}

   As shall be discussed, our method permits us to compute the free energy 
for systems containing a fixed number of vortices at predetermined positions.  
By computing the London energy for different vortex configurations, we
calculate the pinning potential for the vortices. Pinning potentials computed for 
different antidot radii indicate that pinning centers with radii larger 
than $\xi(T)$ are more efficient at pinning than those whose size is in 
the vicinity of the coherence length. This result is in agreement with 
one of the experiments\cite{MoshB98} as well as theoretical work\cite{Takezawa}
on three dimensional superconductors containing columnar defects. The
pinning potentials also allow one to identify interstitial pinning.
With the aid of pinning potentials, we have produced a phase plot 
displaying the optimum number of pinned vortices as a function of defect radius. 
A salient feature is the weak dependence of the number of pinned quanta on the 
effective magnetic penetration depth, $\lambda_{eff}$.  
Finally, we have examined the pinning potentials of defects with an elongated 
shape.  To consider defects with sharp corners, we have 
studied antidots in the shape of a rhombus.  Two notable features 
of the pinning potentials of rhombic defects are a lower pinning 
energy and a higher escape barrier than that of circular defects
of comparable area, qualities indicating that the rhombus is superior 
to the circle as a pinning center.  

   The article is organized as follows:  In Section II, we present 
the wire network model for the supercurrents in the nanoscale defect 
array.  Section III outlines our methods for calculating the London free 
energies and current configurations, and Section IV reports our primary results.
We conclude in Section V with a summary.    

\section{The model}
\begin{figure}
\begin{center}
\centerline{\psfig{figure=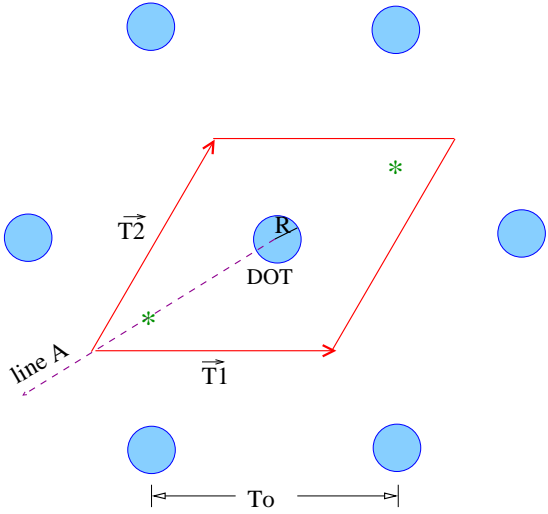,width=3.0in}}
\caption{The triangular lattice with lattice constant $T_{0}$. $\vec{T}_{1}$ and $\vec{T}_{2}$
	are the primitive lattice vectors. $\vec{}T_{1}=T_{0}\hat{x}$, 
	$\vec{T}_{2}=\frac{1}{2}T_{0}\hat{x}+\frac{\sqrt{3}}{2}T_{0}\hat{y}$. 
	The asterisks indicate the centers of the triangular plaquettes defined by the dots.} 
\label{Fig:fig1}        
\end{center}
\end{figure}

To proceed, we consider antidots of radius R forming a regular triangular lattice 
in a thin-film superconductor as shown in Fig.~\ref{Fig:fig1}. The distance between dots is 
$T_0$.  

Since the London theory cannot be solved in closed form for our system, we discretize 
the space of the unit cell including the region of the dot by dividing each edge of the 
unit cell into $N$ intervals. The locations of the $N \times N$ {\em grid} points are 
given by  

\beq
\vec{X}_{n_1,n_2}=\frac{n_1}{N}\vec{T}_{1}+\frac{n_2}{N}\vec{T}_{2}, \qquad  
	0\le n_1, n_2<N.  
\label{Eq:eq1}  
\eeq

The corresponding reciprocal lattice vectors are 
\beq
\vec{G}_{m_1,m_2}=m_1\vec{W}_{1}+m_2\vec{W}_{2},  \qquad  0\le m_1, m_2<N,   
\label{Eq:eq2}  
\eeq

where $\vec{W_{1}}$ and $\vec{W_{2}}$ are the primitive vectors of the reciprocal lattice.

Having discretized the space and defined the grid points, we now construct the current 
network of the unit cell. First, we define 

\begin{itemize}
\item {\em Nodes}: located at the center of each triangular plaquette formed by the three 
		   closest grid points. These nodes form a honeycomb lattice.

\item {\em Links}: are the edges of the honeycomb lattice. The link midpoints
                   are located at the intersection of the links and the plaquette
                   edges.
\end{itemize}

\begin{figure}
\begin{center}
\centerline{\psfig{figure=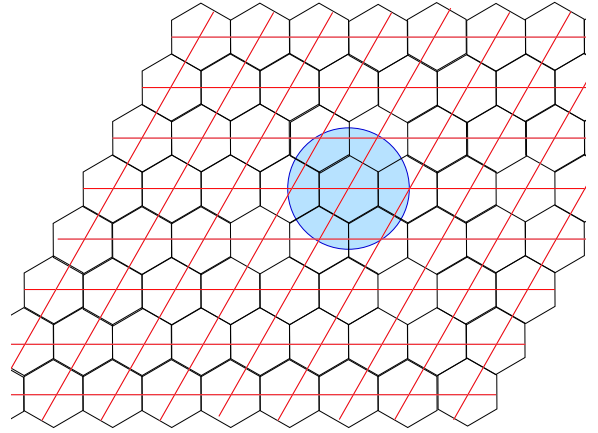,width=3.0in}}
\vspace{5mm}
\caption{The links form a honeycomb lattice, and the current only flows on links. }         
\label{Fig:fig2}
\end{center}
\end{figure}

\begin{figure}
\begin{center}
\centerline{\psfig{figure=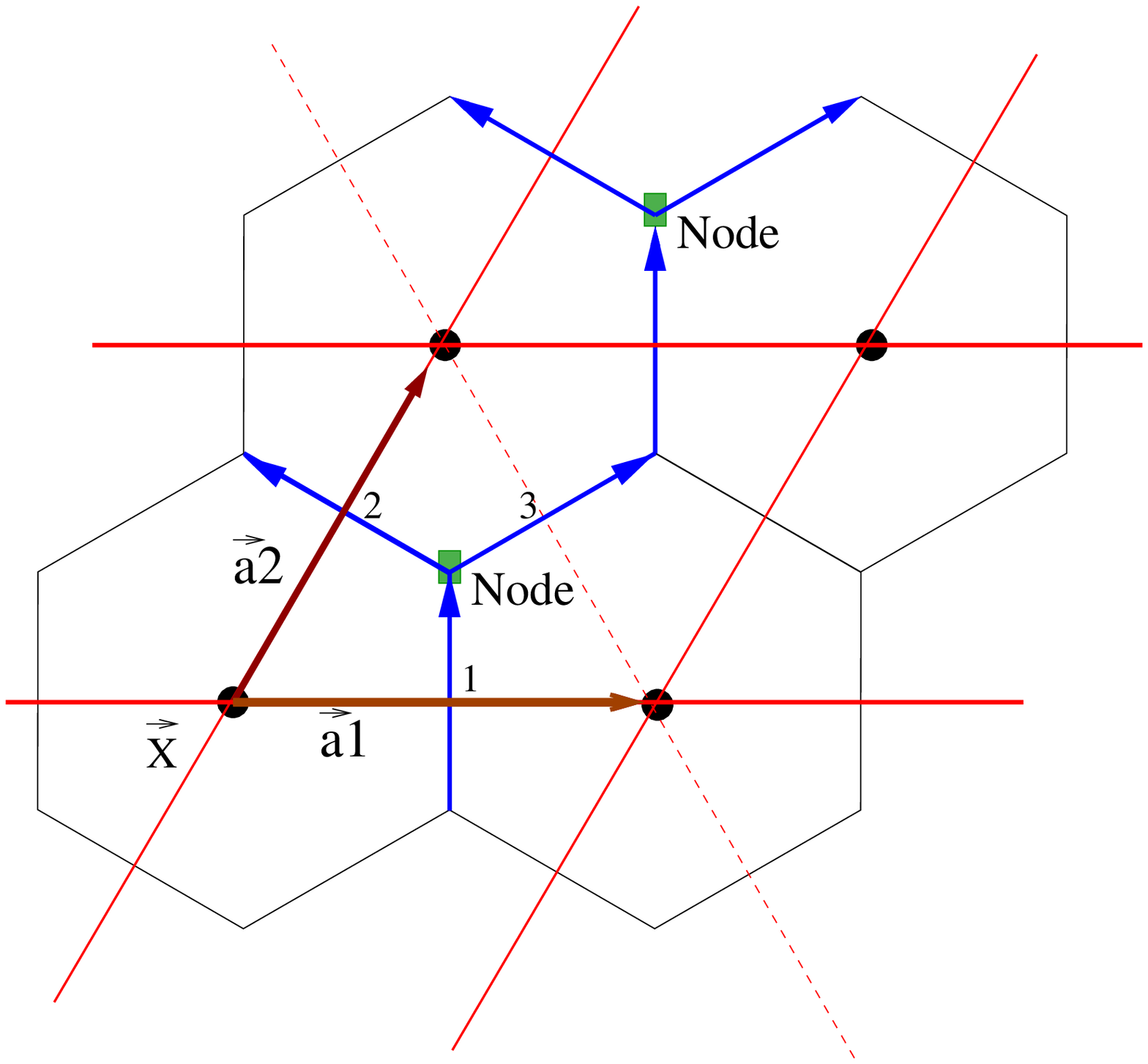,width=3.5in}}
\caption{Currents are depicted for the point $\vec{X}$.  Current conservation 
         requires that $\vec{J_{s1}}=\vec{J_{s2}}+\vec{J_{s3}}$. }         
\label{Fig:fig3}
\end{center}
\end{figure}

We model our system as a wire network. Hence, the currents $\vec{J}_s(\vec{X})$ only flow
on the links. [See Fig.~\ref{Fig:fig2} and Fig.~\ref{Fig:fig3}.]

Fig.~\ref{Fig:fig3} illustrates a detailed section of our wire network. Three current links 
are associated with each grid point  $\vec{X}$, which may be written as

\beq
\vec{J}_s(\vec{X}) = \left[
\begin{array}{c}
       \vec{J}_{s1}(\vec{X})\\
       \vec{J}_{s2}(\vec{X})\\
       \vec{J}_{s3}(\vec{X})
\end{array}
\right],
\label{Eq:eq3}
\eeq

where $\vec{J}_{s1}(\vec{X})$ is centered at $\vec{X}+\frac{1}{2}\vec{a}_1$, 
$\vec{J}_{s2}(\vec{X})$ at $\vec{X}+\frac{1}{2}\vec{a}_2$ and $\vec{J}_{s3}(\vec{X})$ at 
$\vec{X}+\frac{1}{2}(\vec{a}_1+\vec{a}_2)$,
with $\vec{a}_1=\frac{1}{N} \vec{T}_1$ and $\vec{a}_2=\frac{1}{N} \vec{T}_2$.

We now enforce current conservation at the nodes. This gives two equations per grid
cell (i.e., the cell formed by $\vec{a_1}$ and $\vec{a_2}$). 

\beq
\vec{J}_{s1}(\vec{X})=\vec{J}_{s2}(\vec{X})+\vec{J}_{s3}(\vec{X})
\label{Eq:eq4}
\eeq

and

\beq
\vec{J}_{s3}(\vec{X})+\vec{J}_{s2}(\vec{X}+\vec{a}_1)=\vec{J}_{s1}(\vec{X}+\vec{a}_2).
\label{Eq:eq5}
\eeq

After Fourier transforming Eq.(\ref{Eq:eq4}) and (\ref{Eq:eq5}), we find that the current  
can be written in Fourier space as
 
\beq
\vec{J_s}(\vec{G}) = \left[
\begin{array}{c}
       1-e^{-i\vec{G} \cdot \vec{a}_1} \\
       1-e^{-i\vec{G} \cdot \vec{a}_2} \\
       e^{-i\vec{G} \cdot \vec{a}_2}-e^{-i\vec{G} \cdot \vec{a}_1}
\end{array}
\right]\Phi(\vec{G}),  \qquad  \vec{G}\ne 0.
\label{Eq:eq6} 
\eeq

Since there is no net current flowing in the superconductor, we have to include 
the $\vec{G}=0$ component. Therefore, the current at the grid point 
$\vec{X}$ can be written as

\begin{eqnarray}
\vec{J_s}(\vec{X}) & = & j_{01} \left[
\begin{array}{c}
       1\\
       0\\
       1
\end{array}\right] + j_{02} \left[		
\begin{array}{c}
       1/\sqrt{3}\\
       2/\sqrt{3}\\
       -1/\sqrt{3}
\end{array}\right]  \nonumber\\
& + & \sum_{G\ne0}\left[
\begin{array}{c}
       1-e^{-i\vec{G} \cdot \vec{a}_1} \\
       1-e^{-i\vec{G} \cdot \vec{a}_2} \\
       e^{-i\vec{G} \cdot \vec{a}_2}-e^{-i\vec{G} \cdot \vec{a}_1}
\end{array}\right]\Phi(\vec{G})e^{-i\vec{G}\cdot\vec{X}}.
\label{Eq:eq7}
\end{eqnarray}

We can think of
$$
\left[
\begin{array}{c}
	j_{01}\\
	j_{02}\\
	\Phi(\vec{G}_1)\\
	\Phi(\vec{G}_2)\\
	\vdots\\
	\Phi(\vec{G}_{N-1})
\end{array}\right]  
$$

as a vector, whose entries are determined by the current distribution in the
unit cell. Since we are considering systems with antidots, one must have 
no current flowing in the defects. As will be discussed in greater detail in
the following section, we impose this no-current constraint by expressing the
current as a linear combination of elements of a specially constructed basis.
The elements of this basis are themselves current distributions satisfying
the no-current constraint and are conveniently expressed via the vector
formalism mentioned above. The basis elements span the set of all current
configurations for which current is excluded from the defects.

\section{The method}
Generally speaking, one calculates current distributions by minimizing the London
free energy subject to certain constraints, including the absence of current in 
the defects and a fixed number of vortices per unit cell. This shall be our program 
as well. Therefore, we first write the free energy per unit cell of our superconducting 
system in the London approximation,   
  
\beq
F=\frac{2 \pi \lambda_{eff}}{c^2} \sum_{\vec{X}}|\vec{J}_s(\vec{X})|^2
  +\frac{1}{2c} \sum_{\vec{X}} \vec{J}_s(\vec{X}) \cdot \vec{A}_{tot}(\vec{X}).
\label{Eq:eq8}
\eeq

In a thin-film superconductor, if the thickness is much smaller than the London 
penetration depth (i.e., $d<<\lambda$), the volume current density $\vec{J}$ is nearly 
constant in the thickness, so it is appropriate to write $\vec{J}_s=\vec{J}d$ and define 
$\lambda_{eff}=\lambda^2/d$. 
The first term in Eq.(\ref{Eq:eq8}) is the kinetic energy of the supercurrent, 
and the second term is the magnetic field energy, which is the sum of the contributions 
of the supercurrent itself ($\vec{J}_s(\vec{X}) \cdot \vec{A}_s(\vec{X})$) and the 
interaction between the supercurrent and the external field ($\vec{J}_s(\vec{X}) 
\cdot \vec{A}_{ext}(\vec{X})$). However, symmetry arguments indicate that the 
latter term vanishes in the case of thin-film superconductors.  
Since the currents are localized at the link midpoints, we only need to evaluate the dot
product in Eq.(\ref{Eq:eq8}) at the three midpoints in each grid cell. Hence, the free
energy can be written as 

\beqarr
F&=&\frac{2 \pi \lambda_{eff}}{c^2} \sum_{\vec{X}}
	\sum_{j=1}^3|\vec{J^j_s}(\vec{X})|^2 \nonumber\\
  &+&\frac{1}{2c} \sum_{\vec{X}}\sum_{j=1}^3 
	\vec{J^j_s}(\vec{X}) \cdot \vec{A}^j_{s}(\vec{X}) \nonumber\\
 &=&\frac{2 \pi \lambda_{eff}}{c^2} \sum_{\vec{G}}
	\sum_{j=1}^3|\vec{J^j_s}(\vec{G})|^2 \nonumber\\
  &+&\frac{1}{2c} \sum_{\vec{G}}\sum_{j=1}^3 
	\vec{J^j_s}(\vec{G}) \cdot [\vec{A}^j_{s}(\vec{G})]^*.
\label{Eq:eq9}
\eeqarr

In the rest of this section, we will briefly outline our method for solving the London
theory for our system.

\subsection{Constructing the current basis}
To generate a basis satisfying the no-current condition in the defect 
region, we consider a functional which penalizes currents in this region.  
A simple choice is  

\beq
W[\vec{J_s}]=V_0 \sum_{\vec{X}_0}|\vec{J}_s(\vec{X}_0)|^2,
\label{Eq:eq10}
\eeq

where the $\vec{X}_0$'s are the points inside the dot, and $V_0$ is a constant.  
We want to minimize $W[\vec{J_s}]$ subject to the constraint that the current
in the unit cell is normalized. We can implement this constraint by adding a
Lagrangian multiplier to $W[\vec{J_s}]$. The result is a new functional $\mathcal{L}$, 

\beq
\mathcal{L} = \mathrm{V_0 \sum_{\vec{X}_0}|\vec{J}_s(\vec{X}_0)|^2
	- \Lambda \{ \sum_{\vec{X}}|\vec{J}_s(\vec{X})|^2-1 \}}.
\label{Eq:eq11}
\eeq

If we write the supercurrent as a Fourier sum, plug Eq.(\ref{Eq:eq6}) 
into Eq.(\ref{Eq:eq11}), and carry out the sum in real space, we obtain 
$\mathcal{L}$ is a function of $j_{01}, j_{02}$ and $\Phi(\vec{G})$. 
Reality of $\vec{J_s}(\vec{X})$ implies that $\Phi(\vec{-G})=\Phi^*(\vec{G})$. 
This symmetry reduces sums over $\vec{G}$ to sums over $\vec{G}>0$. Next, we write 
$\Phi(\vec{G})=\Phi_{re}(\vec{G})+i\Phi_{im}(\vec{G})$ and minimize $\mathcal{L}$ 
with respect to $j_{01}, j_{02}, \Phi_{re}(\vec{G})$ and $\Phi_{im}(\vec{G})$. This gives 
us four linear equations, which can be written as a matrix equation as Eq.(\ref{Eq:eq12}) 
to solve for $j_{01}, j_{02}, \Phi_{re}(\vec{G})$ and $\Phi_{im}(\vec{G})$.

\beqarr
M[\vec{G}, \vec{G'}]\left[
\begin{array}{c}
	j_{01}\\
	j_{02}\\
	\Phi_{re}(\vec{G'}_1)\\
 	\vdots\\
	\Phi_{re}(\vec{G'}_{(N^2-1)/2})\\
	\Phi_{im}(\vec{G'}_1)\\
 	\vdots\\
	\Phi_{im}(\vec{G'}_{(N^2-1)/2})\\
\end{array}\right] \nonumber\\
=\epsilon \left[
\begin{array}{c}
	j_{01}\\
	j_{02}\\
	\Phi_{re}(\vec{G}_1)\\
 	\vdots\\
	\Phi_{re}(\vec{G}_{(N^2-1)/2})\\
	\Phi_{im}(\vec{G}_1)\\
 	\vdots\\
	\Phi_{im}(\vec{G}_{(N^2-1)/2})\\
\end{array}\right],
\label{Eq:eq12}
\eeqarr

where $M[\vec{G}, \vec{G'}]$ is an $(N^2+1) \times (N^2+1)$ matrix, and $\epsilon \propto
\Lambda/V_0$. 

\begin{figure}
\begin{center}
\centerline{\psfig{figure=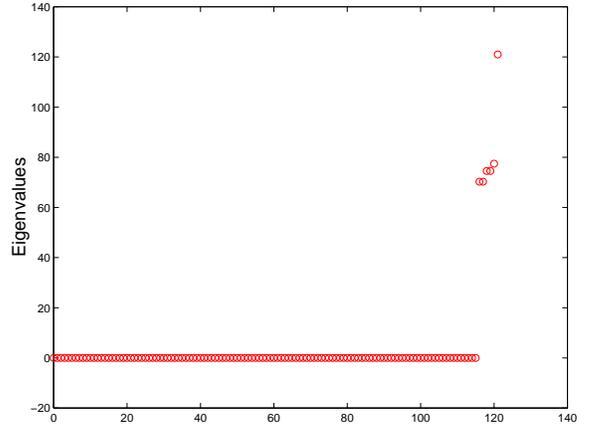,width=3.0in}}
\vspace{5mm}
\caption{The 122 eigenvalues for $N=11$. There are six high values, corresponding to 
	eigenvectors for which current is nonzero in the defect region.
	$T_0=600 nm, R=55 nm$. }
\label{Fig:fig4}
\end{center}
\end{figure}

    After solving the eigenequation, we obtain $(N^2+1)$ eigenvalues. The eigenvalues 
exhibit a clear hierarchy. Fig.~\ref{Fig:fig4} is an example for $N=11$ with $R=55nm$. 
$q$ values are very high relative to the other $N^2+1-q$ eigenvalues. The large 
eigenvalues correspond to eigenstates for which currents within the defect are nonzero. 
Excluding these eigenstates allows us to construct a basis satisfying the no-current condition.
This can be done by inserting the entries of the eigenvector into Eq.(\ref{Eq:eq7})., yielding
$3 \times (N^2+1-q)$ real space versions of the basis elements $\vec{J}^j_n(\vec{X})$ for 
$n=1, 2,...,(N^2+1-q)$, and $j=1, 2, 3$. Any component of the current distribution can 
be expressed as a linear combination of the basis elements, 
i.e., $\vec{J}^j(\vec{X})=\sum_n v_n \vec{J}^j_n(\vec{X})$.

\subsection {Calculating the vector potential}
Having constructed the current basis, we can calculate the vector potential associated with
each element of the current basis. In the discretized scheme, the vector potential is

\beq
\vec{A}(\vec{X})=\frac{1}{c} \sum^{all space}_{\vec{X'}} 
	\frac{\vec{J}(\vec{X'})}{|\vec{X}-\vec{X'}|}.
\label{Eq:eq13}
\eeq

In evaluating Eq.(\ref{Eq:eq13}), one must include the contributions 
from all three types of current. Using the fact that 
$\vec{J}(\vec{X}+\vec{\Re})=\vec{J}(\vec{X})$, where $\vec{\Re}$ is a lattice translation 
vector, we can write

\beqarr
\vec{A}^j(\vec{X})&=&\frac{1}{c} \sum^3_{j'=1}\sum^{unit cell}_{\vec{X'}}\sum_{\vec{\Re}} 
	\frac{\vec{J}^{j'}(\vec{X'})}
	{|\vec{X}-\vec{X'}+\vec{\omega}_j-\vec{\omega}_{j'}-{\vec{\Re}}|} \nonumber\\
	&   &\times \quad \{1-\delta_{\vec{\Re},0}\delta_{\vec{X}, \vec{X'}}\delta_{j, j'}\}.
\label{Eq:eq14}
\eeqarr

$\vec{\omega}_j$ is the displacement from the grid point $\vec{X}$ to link $j=1,2,3$ inside
the grid cell.  To evaluate the slowly converging sum in Eq.(\ref{Eq:eq14}), 
we use the Ewald trick\cite{Ewald}. 
Defining $\xi \equiv |\vec{X}-\vec{X'}+\vec{\omega}_j-\vec{\omega}_{j'}-{\vec{\Re}}|$, 
one uses the identity

\beqarr
\frac{1}{\xi} & = &\frac{2}{\sqrt\pi} \int_{0}^\infty dt e^{-t^2 \xi^2} \nonumber\\
& = & \frac{2}{\sqrt\pi} \int_{0}^\eta dt e^{-t^2 \xi^2}
+\frac{2}{\sqrt\pi} \int_{\eta}^\infty dt e^{-t^2 \xi^2} \nonumber\\
& = & \frac{1}{\xi}erf(\xi\eta)+\frac{1}{\xi}erfc(\xi\eta)
\label{Eq:eq15}
\eeqarr

to break up the sum into a long-ranged piece (the first integral) and a short-ranged 
part (the second integral). The short-ranged term falls off rapidly for $\xi>>1/\eta$,
and is readily computed. 
On the other hand, the long-ranged term converges slowly in real space. 
However, with the aid of two-dimensional form of Ewald's generalized theta function 
transformation \cite{Ewald,Bonsall}

\beqarr
\int d^2\vec{X} e^{-t|\vec{X}+\vec{\omega}_j-\vec{\omega}_{j'}|^2}
		e^{i(\vec{G}+\vec{K}) \cdot \vec{X}} \nonumber\\
=\frac{\pi}{t}e^{-i(\vec{G}+\vec{K}) \cdot (\vec{\omega}_j-\vec{\omega}_{j'})}
	      e^{\frac{-|\vec{G}+\vec{K}|^2}{4t}},
\label{Eq:eq16}
\eeqarr	 

the long-ranged term can be recast in a form that falls off rapidly with increasing $\vec{K}$ 
in reciprocal space, making the sum tractable. (In this context, 
$\vec{K}=N(u_1\vec{W}_1+u_2\vec{W}_2)$, where $u_1$ and $u_2$ are arbitrary
integers.)  The expression for $\vec{A}^j(\vec{X})$, formed by combining the
short and long ranged pieces of the sum, converges reasonably quickly for an 
optimal choice of $\eta$.      
The total vector potential in the system due to  supercurrents at link $j$ of 
$\vec{X}$ is the linear combination of $\vec{A}^j_n(\vec{X})$, i.e., 
$\vec{A}^j(\vec{X})=\sum_n v_n \vec{A}^j_n(\vec{X})$, where $\vec{A}^j_n(\vec{X})$ is the
vector potential associated with the $n$th element of the current basis 
$\vec{J}^j_n(\vec{X})$ via Eq.(\ref{Eq:eq14}).

\subsection {Minimizing the free energy in the presence of constraints}

The Ginzburg-Landau theory yields a set of conditions that allow us to compute the currents
and fields. We will see that these quantities minimize the London free energy while satisfying
system constraints. In the limit that London theory is quantitatively valid, the current
has the form as 

\beq
\vec{J}(\vec{r}) = \frac{e^*n_s}{2m^*} \{ \hbar \vec{\nabla}\varphi(\vec{r}) 
	-\frac{e^*}{c}\vec{A}(\vec{r}) \},
\label{Eq:eq17}
\eeq

where $e^*$ is the superconducting electron charge, $m^*$ the mass, and $n_s$ the number of 
superconducting electrons per $cm^3$. $\varphi(\vec{r})$ is the phase of the order parameter, 
and is therefore a single-valued function. Singularities in $\varphi(\vec{r})$ at $\vec{r}_i$ 
correspond to vortices:

\beq
\nabla^2 \varphi(\vec{r}) = \sum_{i} 2 \pi \delta(\vec{r}-\vec{r}_i)p_i,
\label{Eq:eq18}
\eeq

where $p_i$ is the winding number for a vortex at $\vec{r}_i$. Using the expression for 
 $\vec{J}(\vec{r})$ given in Eq.(\ref{Eq:eq17}), we find 

\beqarr
\oint \vec{J}(\vec{r}) \cdot d\vec{l} & = & 
	\frac{e^*n_s}{2m^*} \{ \hbar \oint \vec{\nabla}\varphi(\vec{r})\cdot d\vec{l}
	- \frac{e^*}{c} \phi_0 \}  \nonumber\\
& = & \frac{\hbar c^2}{4\pi e^* \lambda^2} \{ \sum_{i} p_i - \frac{\Phi_B}{\phi_0} + P\},
\label{Eq:eq19}
\eeqarr

where $\phi_0 (=c\hbar/e^*)$ is superconducting magnetic flux quantum, and $\Phi_B$ 
the magnetic flux contained in path. $P$ is an integer which may arise in 
$\oint \vec{\nabla}\varphi(\vec{r})\cdot d\vec{l}$ for geometries in which the 
superconductor is not simply connected. For example, integrating around a dot can give 
$\oint \vec{\nabla}\varphi(\vec{r})\cdot d\vec{l} \ne 0$ without any singularities. 
Minimizing the London free energy yields

\beqarr
\vec{h}(\vec{r})+\lambda^2 \vec{\nabla} \times\vec{\nabla}\times\vec{h}(\vec{r}) \nonumber\\
=  \vec{h}(\vec{r})+\frac{4\pi \lambda^2}{c} \vec{\nabla} \times\vec{J}(\vec{r})=0.
\label{Eq:eq20}
\eeqarr

Hence, for any path that does not enclose a net vortex charge, we have 

\beq
\oint \vec{J}(\vec{r}) \cdot d\vec{l}  = 
	\frac{c}{4 \pi \lambda^2}\oint \vec{A}(\vec{r}) \cdot d\vec{l},
\label{Eq:eq21}
\eeq

precisely as given by Eq.(\ref{Eq:eq19}). Thus, it is sufficient to seek currents 
satisfying Eq.(\ref{Eq:eq19}). By symmetry, the integral  
$\oint_{cell} \vec{J}(\vec{r}) \cdot d\vec{l}$ vanishes if the integration contour is 
taken to be the unit cell boundary. From this condition, we see that 

\beq
\frac{\Phi_{uc}}{\phi_0}=\sum_{i} p_i+P, 
\label{Eq:eq22}
\eeq

where $\Phi_{uc}$ is the flux through the unit cell. Eq.(\ref{Eq:eq22}) tells us that the 
magnetic field can only have the same periodicity as the dots if the number of flux quanta 
through a unit cell is an integer. For a thin-film superconductor, it is useful to write 
Eq.(\ref{Eq:eq19}) in terms of the surface current density $\vec{J_s}$ and $\lambda_{eff}$:

\beqarr
\oint \vec{J_s}(\vec{X}) \cdot d\vec{l} = 
  \frac{\hbar c^2}{4\pi e^* \lambda_{eff}} \{ \sum_{i} p_i - \frac{\Phi_B}{\phi_0} + P\}.
\label{Eq:eq23}
\eeqarr

Eq.(\ref{Eq:eq23}) gives us two constraints to enforce:

\beq
\oint_{Dot} \vec{J_s}(\vec{X}) \cdot d\vec{l} = 
  \frac{\hbar c^2}{4\pi e^* \lambda_{eff}} \{ P- \frac{\Phi_{Dot}}{\phi_0} \}.
\label{Eq:eq24}
\eeq

and

\beq
\oint_{Sing} \vec{J_s}(\vec{X}) \cdot d\vec{l} = 
  \frac{\hbar c^2}{4\pi e^* \lambda_{eff}} \{ \sum_{i} p_i - \frac{\Phi_{Sing}}{\phi_0} \}.
\label{Eq:eq25}
\eeq

\begin{figure}
\begin{center}
\centerline{\psfig{figure=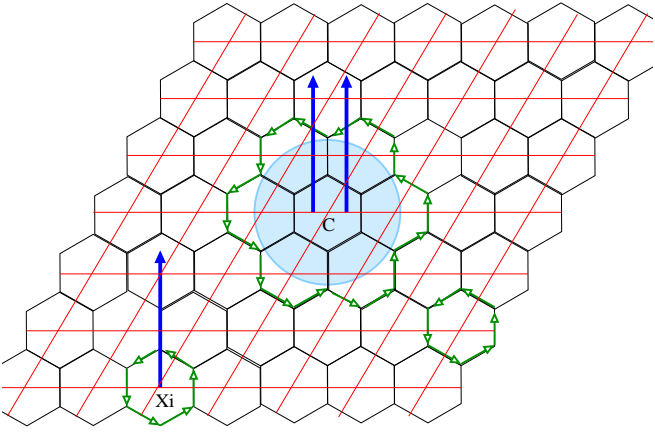,width=3.4in}}
\vspace{5mm}
\caption{The line integral paths around the dot located at $\vec{C}$ and a singularity
        located at $\vec{X}_i$. Paths live on links. The border of the path around the dot  
        consists of the 18 links making up the border of the defect.  
	The path around the singularity is simply the border of the hexagonal plaquette 
        centered at the singularity.} 
\label{Fig:fig5}
\end{center}
\end{figure}

Within our network model, $\oint\vec{J_s}(\vec{X}) \cdot d\vec{l}$ takes the form
of a discrete sum which, when evaluated, depends only on the basis expansion 
coefficients $v_{n}$.  The path of integration is the set of links making up the 
boundary of the defect, and $\oint_{Dot}\vec{J_s}(\vec{X}) \cdot d\vec{l}$ is the sum 
of the inner product of the currents $\vec{J}^j_s$ on the links and the links $\vec{l}_0$. 
In the case of Eq.(\ref{Eq:eq25}), the integral about a singularity, the path consists of
the six edges of the hexagon centered at $\vec{X}$, and 
$\oint_{Sing}\vec{J_s}(\vec{X}) \cdot d\vec{l}$ is
the sum of the six inner products $\vec{J}^j_s \cdot \vec{l}_0$. Fig.~\ref{Fig:fig5} 
depicts an example of the path of integration in the discrete scheme. In the 
figure, the path of integration about the defect contains 18 links, while paths about 
singularities are hexagonal, containing 6 links.

$\Phi_B$, the magnetic flux enclosed by the path of integration, is 

\beqarr
\Phi_B & = & \int d^2\vec{X} \cdot (\vec{h}_{ext}(\vec{X})+\vec{h}_s(\vec{X})) \nonumber\\
       & = & \int d^2\vec{X} \cdot \vec{h}_{ext}(\vec{X})+
	\oint d \vec{l} \cdot \vec{A}_s(\vec{X}). 
\label{Eq:eq26}
\eeqarr

The vector potential $\vec{A}_s(\vec{X})$ is calculated via Ewald method and is
written in terms of $\vec{J}_s(\vec{X})$. In a manner similar to that described 
above for the current line integral, one can compute the line integral  
$\oint d \vec{l} \cdot \vec{A}_s(\vec{X})$.  Again, the result depends only on 
the coefficients $v_{n}$. It is easy to calculate the flux due to the external magnetic 
field. One readily solves Eq.(\ref{Eq:eq22}) for the applied field $h_{ext}$, finding  

\beq
h_{ext}=\frac{\phi_0}{\alpha_{uc}}(\sum_{i} p_i+P),
\label{Eq:eq27}
\eeq

where $\alpha_{uc}$ is the area of the unit cell. Hence, we have for the external
magnetic flux,

\beq
\int d^2\vec{X} \cdot \vec{h}_{ext}(\vec{X})
	=\frac{\alpha_{path}}{\alpha_{uc}}\phi_0(\sum_{i} p_i+P),
\label{Eq:eq28}
\eeq

where $\alpha_{path}$ is the area enclosed by the line of integration. 
Inserting the right side of Eq.(\ref{Eq:eq28}) into Eq.(\ref{Eq:eq24}) and 
performing the discrete version of the line integral yields
 
\beqarr
l_0\sum^{N^2+1-q}_n v_n \sum^3_{j=1}  \sum_{\vec{G} \ne 0} e^{-i\vec{G}\cdot\vec{C}} 
[ \gamma^j(\vec{G}) J^j_n(\vec{G}) ]      \nonumber\\
  = \frac{\hbar c^2}{4 \pi e^* \lambda_{eff}}(P-N_0\frac{\sum_i p_i+P}{N^2}). 
\label{Eq:eq29}
\eeqarr

$\vec{C}$ locates the center of the defect. $\gamma^j(\vec{G})$ is the   
sum of  $\vec{J}^j_s \cdot \vec{l}_0$, and varies with the paths. $N_0$ is 
the number of plaquettes enclosed by the path around the dot.  

Similarly, the discrete version of Eq.(\ref{Eq:eq25}) is

\beqarr
l_0\sum^{N^2+1-q}_n v_n\sum^3_{j=1}  \sum_{\vec{G} \ne 0} e^{-i\vec{G}\cdot\vec{X}_i} 
[ \eta^j(\vec{G}) J^j_n(\vec{G}) ]      \nonumber\\
  = \frac{\hbar c^2}{4 \pi e^* \lambda_{eff}}(p_i\delta_{\vec{X},\vec{X_i}}
-\frac{\sum_i p_i+P}{N^2}), 
\label{Eq:eq30}
\eeqarr

where $\eta^1 \equiv 1-e^{i\vec{G} \cdot \vec{a_1}}$, $\eta^2 \equiv 1-e^{i\vec{G} \cdot \vec{a_2}}$, 
and $\eta^3 \equiv e^{i\vec{G} \cdot \vec{a_2}}-e^{i\vec{G} \cdot \vec{a_1}}$.

Note that Eq.(\ref{Eq:eq30}) holds for each of the plaquettes external to the dot, 
yielding a total of $N^2-(q+1)$ equations. Equations(\ref{Eq:eq29}) and (\ref{Eq:eq30}) 
allow us to solve uniquely for the coefficients $v_{n}$, permitting  one to calculate the 
currents corresponding to a particular vortex configuration.

\subsection {Solving for $v_n$'s}

   Since the current in the system is given by $\vec{J}^j_s(\vec{X})=\sum_n v_n
\vec{J}^j_n(\vec{X})$, solving Equations(\ref{Eq:eq29}) and (\ref{Eq:eq30}) for 
the coefficients $v_{n}$ is tantamount to finding the current distribution.  
As previously mentioned, we seek $N^2+1-q$ of the coefficients $v_{n}$, where 
$q+1$ is the number of plaquettes comprising the defect. Eq.(\ref{Eq:eq29}), and
Eq.(\ref{Eq:eq30}) represent a total of $N^2-q$ equations. However, it may be shown
that only $N^2-(q+1)$ of these are independent.

Requiring that the net supercurrent vanish yields

\beq
\sum_n v_n j^n_{01}=0
\label{Eq:eq32}
\eeq

and

\beq
\sum_n v_n j^n_{02}=0.
\label{Eq:eq33}
\eeq

One now has the desired $N^2+1-q$ independent linear equations for the 
coefficients $v_{n}$.  In matrix form, we have

\beqarr
\left[
\begin{array}{ccccc}
a_{11} & a_{12}  & \cdots & a_{1i} & \cdots  \\
a_{21} & a_{22}  & \cdots & a_{2i} & \cdots \\
\vdots & \vdots  & \ddots & \vdots & \ddots  \\
a_{i1} & a_{i2}  & \cdots & a_{ii} & \cdots  \\
\vdots & \vdots & \ddots & \vdots & \ddots  \\
\end{array}
\right]
\left[
\begin{array}{c}
	v_1\\
        v_2\\
	\vdots\\
	v_i\\
	\vdots\\ 
\end{array}
\right]=\left[
\begin{array}{c}
	P-\frac{\sum_i p_i+P}{N^2}\\
        0-\frac{\sum_i p_i+P}{N^2}\\
	\vdots\\
	p_i-\frac{\sum_i p_i+P}{N^2}\\
	\vdots\\ 
\end{array}
\right].
\label{Eq:eq34}
\eeqarr

The coefficients $a_{ij}$ are found from Eqs.(\ref{Eq:eq29})-(\ref{Eq:eq33}).
Solving Eq.(\ref{Eq:eq34}) yields the current distribution and, via Eq.(\ref{Eq:eq9}), 
the free energy.

\section{results}

\subsection{Pinning Potentials}

\begin{figure}
\begin{center}
\centerline{\psfig{figure=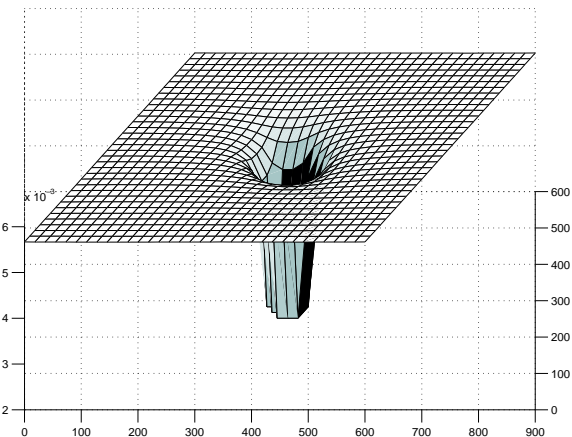,width=3.0in}}
\vspace{5 mm}
\caption{Energy surface for a single vortex in the unit cell of antidots.  
	$T_0=600 nm$, $R=0.06T_0$, $\lambda=1.5T_0$.}         
\label{Fig:fig6}
\end{center}
\end{figure}

The freedom to control the locations of vortices in the unit cell permits
the calculation of pinning potentials.  The simplest case, in which a vortex
interacts with a pinning center containing zero flux quanta, is displayed in 
Fig.~\ref{Fig:fig6}.  One can see that the vortex is strongly pinned by the 
dot because the potential drops discontinuously when the vortex enters the
defect zone.  The potential surface in the vicinity of the dot forms a 
potential well with a length scale set by the defect radius.  

\begin{figure}
\begin{center}
\centerline{\psfig{figure=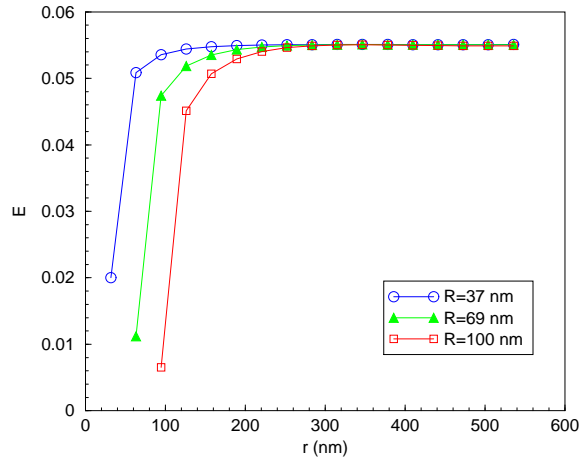,width=3.0in}}
\caption{Pinning Potentials for a single vortex in the unit cell of 
antidots with three different
	radii. $T_0=600 nm$, $\lambda_{eff}=90 nm$. }         
\label{Fig:fig7}
\end{center}
\end{figure}

   Examining a cross section of the pinning potential allows one to infer the 
pinning energy, the energy required to dislodge a single vortex from the 
interior of the defect.  Fig.~\ref{Fig:fig7} displays potential cross sections 
for defect radii spanning the range between the coherence length  
$\xi(T)$ and $\lambda_{eff}$.  A monotonic lowering of the pinning potential 
with increasing dot radius is evident from the plot.  The graph in 
Fig.~\ref{Fig:fig7} supports the conclusion that the pinning energy 
initially increases rather swiftly as the radius of the defect increases 
beyond the coherence length.

\begin{figure}
\begin{center}
\centerline{\psfig{figure=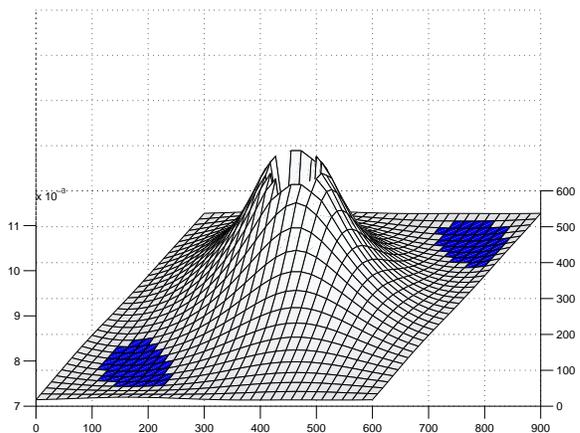,width=3.0in}}
\vspace{5 mm}
\caption{Energy surface in the unit cell of saturated antidots ($n_s=1$). 
	$T_0=600 nm$, $R=0.06T_0$, $\lambda=1.5T_0$.}         
\label{Fig:fig8}
\end{center}
\end{figure}

   Interesting physics arises when one computes the pinning potential for a 
vortex interacting with a defect containing a single flux quantum.  The pinning  
potential for a typical case is shown in Fig.~\ref{Fig:fig8}.  The energy 
barrier surrounding the dot is perhaps the most salient qualitatively 
new feature compared to Fig.~\ref{Fig:fig7}. The barrier tends to repel the 
external vortex from the defect zone.  However if the vortex manages to cross 
the barrier, the pinning potential abruptly decreases and the vortex finds itself in a 
local minimum.  

    As the image indicates, another local minimum occurs at sites 
located at the centers of the equilateral triangles formed by the 
antidots.  This interstitial minimum is qualitatively very similar to 
that discussed in Ref.~\ref{Ref:ref16}.  As the darkened regions in 
Fig.~\ref{Fig:fig8} indicate, the interstitial pinning potential is 
quite shallow.  Therefore, one expects a higher mobility for    
interstitially pinned vortices than for vortices captured by the 
dots.

\begin{figure}
\begin{center}
\centerline{\psfig{figure=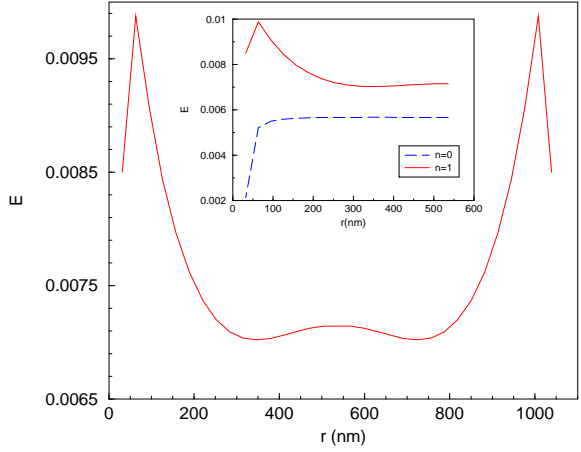,width=3.0in}}
\vspace{3 mm}
\caption{Energy of vortex vs. the distance from the center of the antidot along line A in 
	Fig.~\ref{Fig:fig1} when each dot is saturated with one vortex. The inset shows both 
	curves for empty and saturated dots.}         
\label{Fig:fig9}
\end{center}
\end{figure}

Fig.~\ref{Fig:fig9} displays a cross section of the single quantum pinning potential.
The cross section is calculated along line A as illustrated in Fig.~\ref{Fig:fig1}.
The domain in the graph spans the distance between adjacent defects.  Evidently,
the presence of an energy barrier surrounding the dot necessitates the existence 
of an interstitial minimum.  The inset offers a comparison with the case in which 
the pinning centers are not occupied by vortices.  The pinning potential cross
sections are qualitatively very similar to potentials calculated in
Ref.~\ref{Ref:ref6} and Ref.~\ref{Ref:ref11}.

\begin{figure}
\begin{center}
\centerline{\psfig{figure=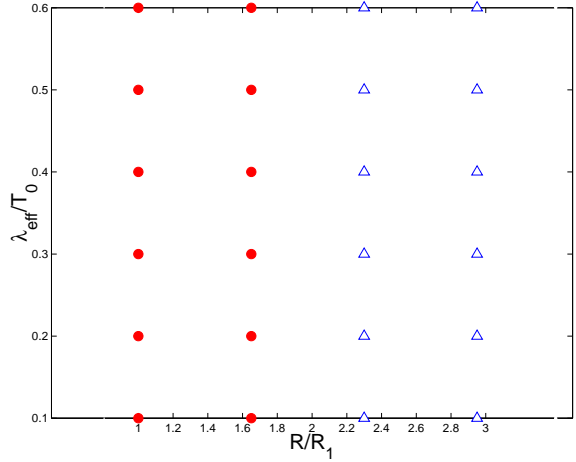,width=3.0in}}
\vspace{5 mm}
\caption{Pinning phase diagram with weak dependence on $\lambda_{eff}$.
The close-face squares 
	 represent the case of one captured vortex and one interstitial vortex,
while the open-face triangles indicate that two vortices are captured.
	 $T_0=600 nm, R_1=0.05T_0$. }         
\label{Fig:fig10}
\end{center}
\end{figure}

Fig.~\ref{Fig:fig10} displays a pinning phase diagram.  In the diagram the 
saturation number $n_{s}$, the number of flux quanta captured by the dot, 
is shown as a function of defect radius $R$ and $\lambda_{eff}$.  The phase diagram
is plotted for fixed $\xi(T)$.  Since the graininess of the discretization 
makes the notion of ``radius'' ambiguous, we take the defect radius to 
be the radius of the circle whose area is the same as the area enclosed
by the defect boundary.  A salient feature of the phase diagram is the absence
of a strong dependence on the effective penetration depth, $\lambda_{eff}$\cite{work}.  
In fact, the chief factor in the pinning phenomena for fixed $\xi(T)$ 
seems to be the defect radius.  The dependence of $n_{s}$ on $R$ fits one's 
intuition.  For small values of $R$, it is energetically favorable for the 
dots to capture one vortex each, while the remaining vortex resides at the
interstitial site.  For larger defects, the two vortex capture scenario 
becomes increasingly favorable.  When the dot is made larger than about $0.15$
lattice constant, the trapping of two quanta prevails and $n_{s} = 2$.

\subsection{Pinning potentials for noncircular defects}

\begin{figure}
\begin{center}
\centerline{\psfig{figure=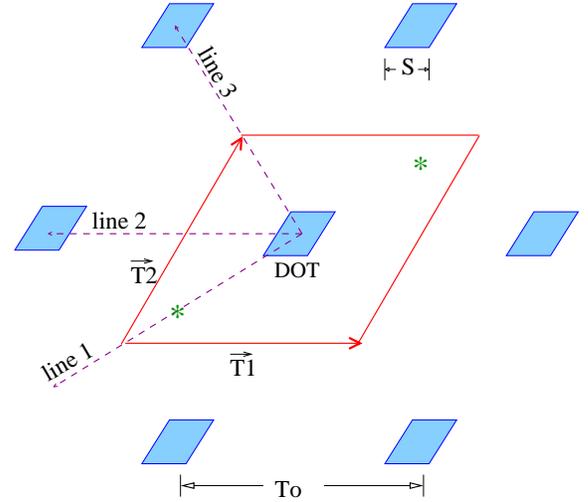,width=3.0in}}
\caption{The triangular lattice of antidots in the shape of rhombus. Each side of the dot
	 has length $S$.}         
\label{Fig:fig11}
\end{center}
\end{figure}

\begin{figure}
\begin{center}
\centerline{\psfig{figure=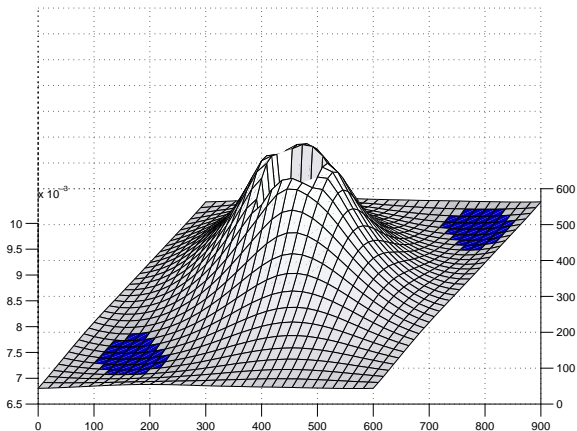,width=3.0in}}
\vspace{5 mm}
\caption{Energy surface for a single vortex in the unit cell of saturated rhombic
defects containing one vortex each
	($n_s=1$). $T_0=600 nm$, $S \sim 0.16T_0$, $\lambda=1.5T_0$.}         
\label{Fig:fig12}
\end{center}
\end{figure}

   It is interesting to examine pinning phenomena for cases in which 
the pinning center has an elongated structure. With the geometry of the triangular
lattice, it is convenient to study defects in the shape of a rhombus, as illustrated 
in Fig.~\ref{Fig:fig11}.  It is informative to calculate pinning potentials 
in which a vortex interacts with defects containing a single flux quantum.  The 
potential surface is depicted in Fig.~\ref{Fig:fig12}.   
As for the circular case, energy minima appear near the centers of equilateral
triangles formed by the defects.  The structure of the interstitial sites, 
including the shallowness of the associated minima, is very similar to that of the 
circular case.  
   As in the case of circular defects, the occupation of the rhombic defect 
by a vortex is associated with an energy barrier.  However, the barrier in 
the rhombic case is distinguished by a marked anisotropy.  The barrier near the 
acute-angled corners of the rhombus is lower than the barrier height at any other 
part of the defect boundary.  One readily concludes that vortices tend to enter
or leave defects through the sharper points.  This phenomenon is reminiscent of 
the manner in which electrons on a conductor preferentially discharge from the 
region of greatest curvature.

\begin{figure}
\begin{center}
\centerline{\psfig{figure=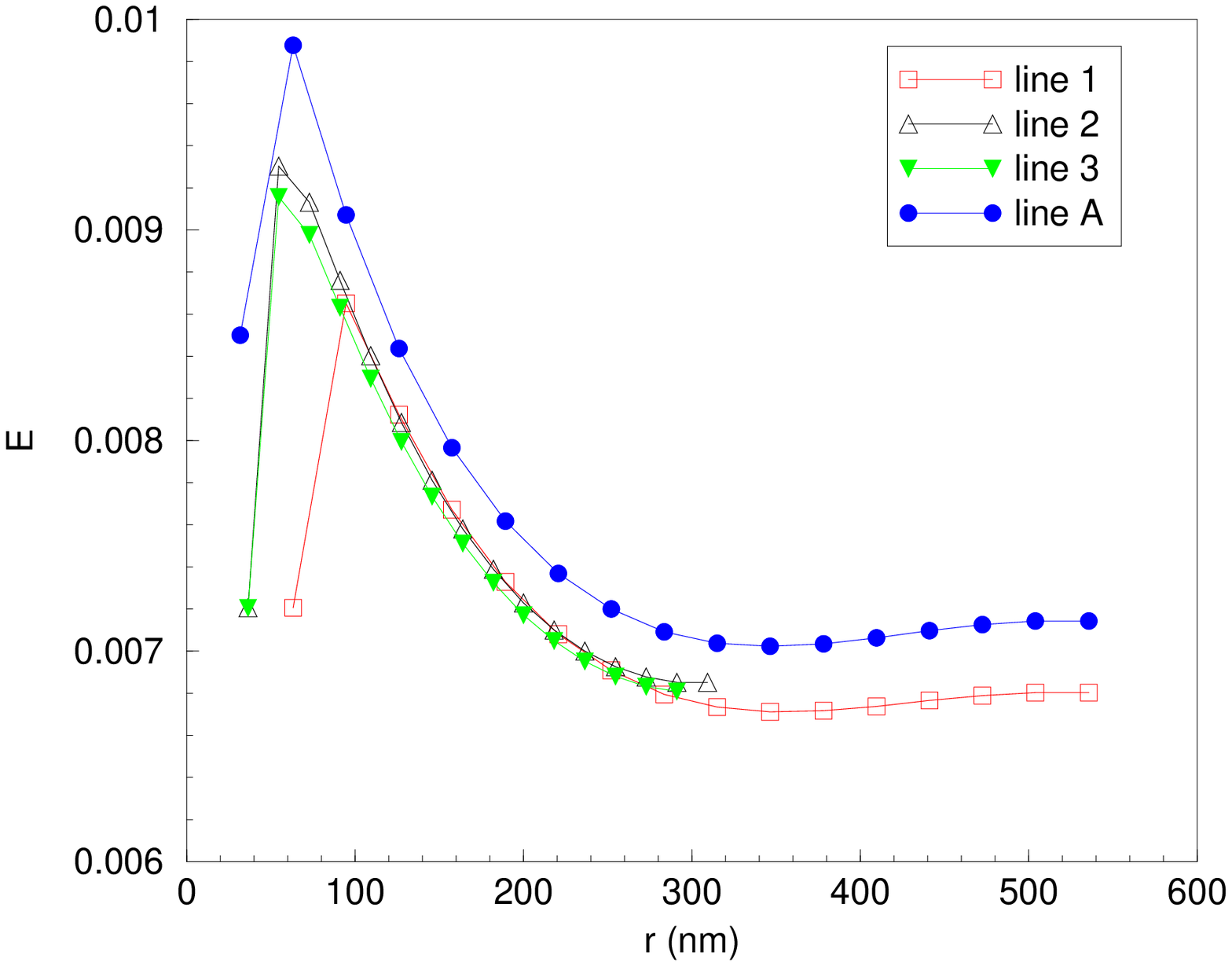,width=3.0in}}
\vspace{5 mm}
\caption{Energy of vortex vs. the distance from the center of the antidot along line A in 
	Fig.~\ref{Fig:fig1}, line 1, line 2 and line 3 in Fig.~\ref{Fig:fig11} when each dot 
	is saturated with one vortex. $T_0=600 nm, \lambda_{eff}=1.5T_0, S=2R \sim 0.16T_0$.}        
\label{Fig:fig13}
\end{center} 
\end{figure}

\begin{figure}
\begin{center}
\centerline{\psfig{figure=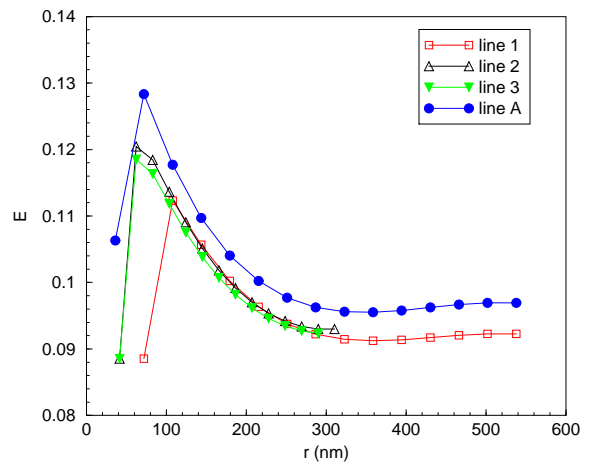,width=3.0in}}
\vspace{5 mm}
\caption{Energy of vortex vs. the distance from the center of the antidot along line A in 
	Fig.~\ref{Fig:fig1}, line 1, line 2 and line 3 in Fig.~\ref{Fig:fig11} when each dot 
	is saturated with one vortex. $T_0=600 nm, \lambda_{eff}=0.1T_0, S=2R \sim 0.16T_0$.}        
\label{Fig:fig14}
\end{center} 
\end{figure}

   It is useful to compare the second capture energies of circular 
and rhombic defects of comparable area.  (In this context, the $N$th  
capture energy is defined to be the shift in energy which occurs when 
a vortex is moved from the interstitial region to the interior of the defect 
to join $N-1$ previously captured quanta.)  To aid in the comparison, we 
display on the same plot pinning potential cross sections for defects with both
circular and rhombic shapes in Fig.~\ref{Fig:fig13} and Fig.~\ref{Fig:fig14}.
The lone circular defect pinning potential is plotted along a radial line 
for the circular case.  The 3 rhombic potentials shown on the same graph correspond 
to paths of approach which the reader can identify from Fig.~\ref{Fig:fig11}.  
``Line 1'' represents the easiest line of approach for the rhombus, in which one 
moves along the symmetry axis of the rhombus directly toward the sharpest corner 
of the defect.  Two features of the potential are favorable for pinning.  
The ``energy of approach'', defined as the energy needed to move the vortex 
to the crest of the energy barrier, is lower for the rhombic defect than for the 
circular dot. The potential also indicates a greater pinning energy for rhombic 
defects. In fact, Fig.~\ref{Fig:fig14} displays a case in which it is favorable for 
a rhombic defect to capture two flux quanta while its circular counterpart
is only able to capture a single vortex.  The lower energy of approach 
ensures faster kinetics, hence swifter vortex capture for elongated pinning 
centers.  The greater pinning energy, on the other hand, ensures that 
the vortices remain trapped once captured.

\section{Summary}

In the work described in this report, we have used a wire network model to 
provide a detailed treatment of pinning phenomena in nanoscale defect arrays.  
Our results are consistent with experiments involving thin-film 
superconductors and are qualitatively very similar to theoretical work 
on three dimensional superconductors containing columnar defects.
 
A comparison of pinning potentials suggests that elongated pinning centers 
have superior pinning characteristics to similar sized circular defects.   

Future work will extend this study to more general defect types and lattice 
geometries.  We have fashioned techniques that permit the study of 
films with nontrivial three dimensional structures. Many of the experiments 
performed on nanoscale defect arrays are neither deep in the thin-film limit 
nor in the bulk limit. To model these systems properly, it is necessary to 
go beyond the thin-film approximation and provide a detailed three dimensional 
treatment of the superconducting films. Such studies will be presented in
future work.

\vspace{15mm}
\centerline{\bf ACKNOWLEDGMENTS}
\vspace{5mm}

The authors would like to thank Professor Lance DeLong for useful discussions 
concerning this research. This work was supported by NSF Grant No. DMR-9870681
and DMR-0108451.


\begin{thebibliography}{99}
\bibitem{work}
It can be shown that some dependence of $n_s$ on $\lambda_{eff}$ does exist. 
These calculations will be presented elsewhere.
\end{thebibliography}
\end{document}